\documentclass[conference]{IEEEtran}
\IEEEoverridecommandlockouts
\usepackage{cite}
\usepackage{amsmath,amssymb,amsfonts}
\usepackage{algorithmic}
\usepackage{graphicx}
\usepackage{textcomp}
\usepackage{xcolor}
\def\BibTeX{{\rm B\kern-.05em{\sc i\kern-.025em b}\kern-.08em
    T\kern-.1667em\lower.7ex\hbox{E}\kern-.125emX}}
\begin{document}

\title{Classifying Peace in Global Media Using RAG and Intergroup Reciprocity\\ \thanks{Funding provided by the Climate School at Columbia University for student researchers and from the Toyota Research Institute.}
}

\author{\IEEEauthorblockN{Kevin Lian}
\IEEEauthorblockA{\textit{Fu Foundation School of} \\  \textit{Engineering and Applied Science} \\
\textit{Columbia University}\\
New York, NY, USA \\
kl3451@columbia.edu}

\and
\IEEEauthorblockN{ Larry S. Liebovitch}
\IEEEauthorblockA{\textit{AC4 in the Climate School} \\
\textit{Columbia University}\\
New York, NY, USA \\
lsl2140@columbia.edu}

\and
\IEEEauthorblockN{Melissa Wild}
\IEEEauthorblockA{\textit{AC4 in the Climate School} \\
\textit{Columbia University}\\
New York, NY, USA \\
mm3484@columbia.edu}

\and
\IEEEauthorblockN{Harry West}
\IEEEauthorblockA{\textit{Indust.l Engr. \& Ops. Research} \\
\textit{Columbia University}\\
New York, NY, USA \\
hw2599@columbia.edu}

\and
\IEEEauthorblockN{Peter T. Coleman}
\IEEEauthorblockA{\textit{Teachers College,}  \\ \textit{AC4 in the Climate School} \\
\textit{Columbia University}\\
New York, NY, USA \\
coleman@exchange.tc.columbia.edu}

\and
\IEEEauthorblockN{Francine Chen}
\IEEEauthorblockA{\textit{Harmonious Communities,} \\ \textit{Human-Centered AI} \\
\textit{Toyota Research Institute}\\
Los Altos, CA, USA \\
francine.chen@tri.global}
\and
\IEEEauthorblockN{Everlyne Kimani}
\IEEEauthorblockA{\textit{Harmonious Communities,} \\ \textit{Human-Centered AI} \\
\textit{Toyota Research Institute}\\
Los Altos, CA, USA \\
everlyne.kimani@tri.global}
\and
\IEEEauthorblockN{Kate Sieck}
\IEEEauthorblockA{\textit{Harmonious Communities,} \\ \textit{Human-Centered AI} \\
\textit{Toyota Research Institute}\\
Los Altos, CA, USA \\
kate.sieck@tri.global}

}

\maketitle

\begin{abstract}
This paper presents a novel approach to identifying insights of peace in global media using a Retrieval-Augmented Generation (RAG) model and concepts of Positive and Negative Intergroup Reciprocity (PIR/NIR). By refining the definitions of PIR and NIR, we offer a more accurate and meaningful analysis of intergroup relations as represented in media articles. Our methodology provides insights into the dynamics that contribute to or detract from peace at a national level.
\end{abstract}

\section{Introduction}

Understanding how peace is represented in global media is crucial for gaining insights into the factors that influence a nation's peace level. Traditional approaches to analyzing media representations have typically relied on word frequency or sentiment analysis, which can overlook the nuanced dynamics of intergroup relations. In contrast, this study introduces a novel methodology that utilizes a Retrieval-Augmented Generation (RAG) model to identify and quantify Positive and Negative Intergroup Reciprocity (PIR/NIR) within media articles. By examining these concepts, we can classify countries based on their alignment with PIR or NIR, providing a more sophisticated understanding of how intergroup relations contribute to or detract from national peace.

The data used in this study was drawn from the News On the Web (NOW) dataset, which contains 700,000 articles comprising 58 million words. The mean number of articles per country across the 18 countries included in the analysis was 40,199, with a standard deviation of 24,214 articles. Similarly, the mean number of words per country was 3,212,191, with a standard deviation of 2,083,456 words. As described by Liebovitch et al. \cite{plosone}
\begin{quote} ``the dataset includes a wide range of topics such as accidents, business, crime, education, the arts, government, healthcare, law, literature, medicine, politics, real estate, religion, sports, war, as well as book, music, and movie reviews. Some examples of media sources included in the dataset are AlterNet, Business Insider, Chicago Tribune, USA TODAY, Jerusalem Post, and Vulture, among many others.''
\end{quote}

\subsection{Background}

Intergroup relations, defined by the dynamics of positive and negative reciprocity between groups, are essential for shaping a nation's peace. Positive Intergroup Reciprocity (PIR) refers to the presence of language in media articles that emphasizes intergroup tolerance, respect, kindness, help, or support. PIR focuses on the portrayal of positive intergroup dynamics where terms associated with cooperation, understanding, and mutual respect are prominent. These terms indicate media coverage that fosters cooperation, social harmony, and peaceful coexistence between groups. Conversely, Negative Intergroup Reciprocity (NIR) refers to language that highlights intergroup intolerance, disrespect, aggression, obstruction, or hindrance. NIR captures the use of negative terms and descriptions that suggest intergroup tensions, conflict, or hostile interactions. This includes media portrayals that amplify discord, exclusion, or aggression between groups, contributing to a narrative of division and conflict. Previous research, including work by Coleman et al. \cite{ampsych}, has highlighted the importance of these intergroup processes in both high and low peace cultures and modeled their dynamics using nonlinear, differential equations \cite{lsl}. However, while the theoretical foundations of PIR and NIR have been well-documented, including in works such as the one by Wang et al. \cite{wang}, there has been a notable absence of methodologies to measure these concepts effectively in real-world scenarios. Our study addresses this gap by utilizing a Retrieval-Augmented Generation (RAG) model, which allows for the practical identification and measurement of PIR and NIR in media articles.

\subsection{Objectives and Contributions}

The primary contributions of this paper are:

\begin{enumerate}
    \item \textbf{Refinement of PIR/NIR Definitions:} We refine the definitions of Positive and Negative Intergroup Reciprocity to enhance the accuracy and relevance of our analysis.
   
    \item \textbf{RAG Implementation for Identifying PIR and NIR:} We use a Retrieval-Augmented Generation model to identify media articles that exemplify PIR and NIR, allowing us to classify countries based on their alignment with these intergroup dynamics.

    \item \textbf{Cross-National Peace Analysis:} By applying our methodology across multiple countries, we provide insights into how intergroup relations, as portrayed in media, impact national peace levels, offering an alternative perspective to traditional peace measurements.
\end{enumerate}

\section{Methods}

This study employs a systematic and multi-faceted approach to classify and analyze media articles using advanced machine learning techniques, including the integration of large language models (LLMs) through a Retrieval Augmented Generation (RAG) process. The methodology is divided into four primary components: embedding the articles, identifying Positive Intergroup Reciprocity (PIR) and Negative Intergroup Reciprocity (NIR) in various countries, implementing LLM queries with the RAG framework, and performing machine learning classification via cosine similarity,. This section provides an in-depth explanation of each component.

\subsection{Embedding the Articles}

The process of embedding articles into high-dimensional vectors is a foundational step that enables subsequent similarity computations and classifications. This process, executed using Python scripts, involves several crucial steps:

\subsubsection{Data Preprocessing}

Data preprocessing is essential for preparing the raw text data to ensure that the embeddings accurately reflect the semantic content of the articles. Initially, we considered applying several traditional preprocessing steps:

\begin{itemize}
    \item \textbf{Tokenization}: Splitting the text into individual tokens or words, which are the basic units of analysis in natural language processing.
    \item \textbf{Stop Words Removal}: Removing common words (such as "and", "the", "is") that are generally uninformative and do not contribute significantly to the overall meaning.
    \item \textbf{Punctuation Removal}: Stripping out punctuation marks, which do not carry semantic content and can introduce noise in the analysis.
    \item \textbf{Stemming/Lemmatization}: Reducing words to their base or root form to standardize different morphological variations (e.g., "running" becomes "run").
\end{itemize}

However, after thorough evaluation, we opted to retain the raw, unfiltered text data. This decision was informed by the capabilities of modern LLMs, which are trained on extensive corpora of natural language and can effectively handle unprocessed text. Retaining the natural language structure allows the model to leverage context and nuanced meanings, potentially enhancing the quality and accuracy of the embeddings generated.

\subsubsection{Embedding Generation}

Following preprocessing, each article is tagged with metadata, including the country of origin and its associated peace level. The high/low peace tags were applied as described in the article by Liebovitch et al. \cite{plosone}. This enriched text is then passed through a pre-trained embedding model, specifically the OpenAI text-embedding-3-small model. This model transforms the text into a 1536-dimensional vector, which captures the semantic essence of the article. These high-dimensional vectors are critical for downstream tasks, such as calculating similarities between articles and classifying their content.

\subsubsection{Storage and Retrieval}

The generated embeddings are stored in a ChromaDB database, which is optimized for handling large volumes of vectorized data. This database architecture supports efficient retrieval and comparison during the classification and querying phases. The storage solution not only accommodates the vast number of articles in the dataset but also ensures quick access to embeddings, facilitating real-time or near-real-time analysis.

\subsection{Retrieval Augmented Generation (RAG) Implementation}

The final and most sophisticated component of our methodology is the Retrieval Augmented Generation (RAG) model. RAG integrates large language models (LLMs) with a retrieval system to generate responses that are contextually informed by both pre-existing social science knowledge and specific media content. The process consists of several key steps:

\subsubsection{User Input and Initial Query}

The RAG process begins when the user provides a query. This query typically relates to a specific social process or theme they want to measure or analyze, such as intergroup tolerance or aggression. The query is then passed into the retrieval system, initiating the search for relevant contextual information.

\subsubsection{Social Science Database Search}

First, the retrieval system searches the Social Science Database, which consists of academic references, particularly those drawn from Douglas Fry’s work on peace and conflict studies \cite{fry}. This database is searched using only the original user query. The purpose of this step is to define and enrich the user’s input by retrieving relevant theoretical or empirical background. This contextual information helps refine the query and grounds the analysis in established social science principles.

\subsubsection{Article Database Search}

Next, the retrieval system searches the Article Database, which contains a large corpus of media articles. In this step, both the original user query and the embeddings of the results returned from the Social Science Database search are used. This combined input allows the system to identify articles that are not only relevant to the original query but also contextually aligned with the theoretical insights provided by the social science references. For example, if the user is analyzing media portrayals of intergroup aggression, the system retrieves articles that discuss instances of aggression, hostility, or related themes, grounded in the context defined by the Social Science Database.

\subsubsection{Contextual Augmentation}

The retrieved documents from the Social Science Database and the Article Database are then combined to augment the initial query. This augmented context provides a richer, more comprehensive understanding of the social process in question, linking theoretical insights from social science literature with practical examples from real-world media content.

\subsubsection{LLM Processing and Response Generation}

The augmented query, now enriched with additional context from both the Social Science and Article Databases, is passed to the large language model (LLM). The LLM processes this input and generates a response based on the integrated information. The response could take various forms, such as identifying patterns within the media, summarizing relevant content, or offering a detailed analysis of the social process. For instance, the LLM might identify underlying social dynamics such as positive or negative intergroup reciprocity (PIR or NIR) from the media articles based on the enriched input.

For example, in peaceful countries like Canada (CA), the LLM output might describe, "The articles highlight the existence of intergroup intolerance and disrespect, particularly through organized groups like Pegida and the Soldiers of Odin, which have increased activity and public demonstrations." This indicates that "even in societies considered peaceful, there is an undercurrent of intolerance that could jeopardize long-term peace if not addressed." Additionally, the LLM output includes relational aggression, noting that "Canada also experiences relational aggression, with boys using social exclusion and spreading rumors, which can hamper social cohesion."

In Singapore (SG), the LLM output may focus on religious appreciation and social harmony: "In Singapore, there is a deliberate effort to move beyond mere tolerance to a deeper appreciation of different religious beliefs. This appreciation fosters unity and a collective pursuit of common good, enhancing intergroup respect and support." Moreover, the LLM output references research that "indicates that social networks and social capital within organizations can improve productivity and cooperation, which supports a harmonious social environment."

\subsubsection{Response Delivery}

Once the LLM generates a response, it is returned to the retrieval system and then delivered to the user. The nature of the response depends on the user’s initial query. It could include identification of relevant social processes, summaries of media portrayals, or more in-depth explanations of the dynamics found in the articles. For instance, if the user requested an analysis of intergroup reciprocity in media, the response may include examples like those from Canada, where "organized groups have increased activity," or Singapore, where efforts to "move beyond mere tolerance" are highlighted, demonstrating both the positive and negative dynamics that contribute to or detract from peace in those countries.

\subsubsection{Advantages of RAG in Peace Studies}

The RAG approach offers several advantages in the field of peace studies. By combining the knowledge of large language models with a retrieval system, RAG allows for a nuanced and contextually informed analysis of complex social processes, such as intergroup relations, peace, and conflict. This method enables the integration of theoretical knowledge with real-time media data, providing insights that are both empirically grounded and relevant to ongoing social dynamics. RAG’s ability to link academic context with real-world content is especially useful for examining multifaceted issues like peace, where the interplay of various social factors must be understood in context.

\begin{figure}[htbp]
    \centering
    \includegraphics[width=0.58\textwidth]{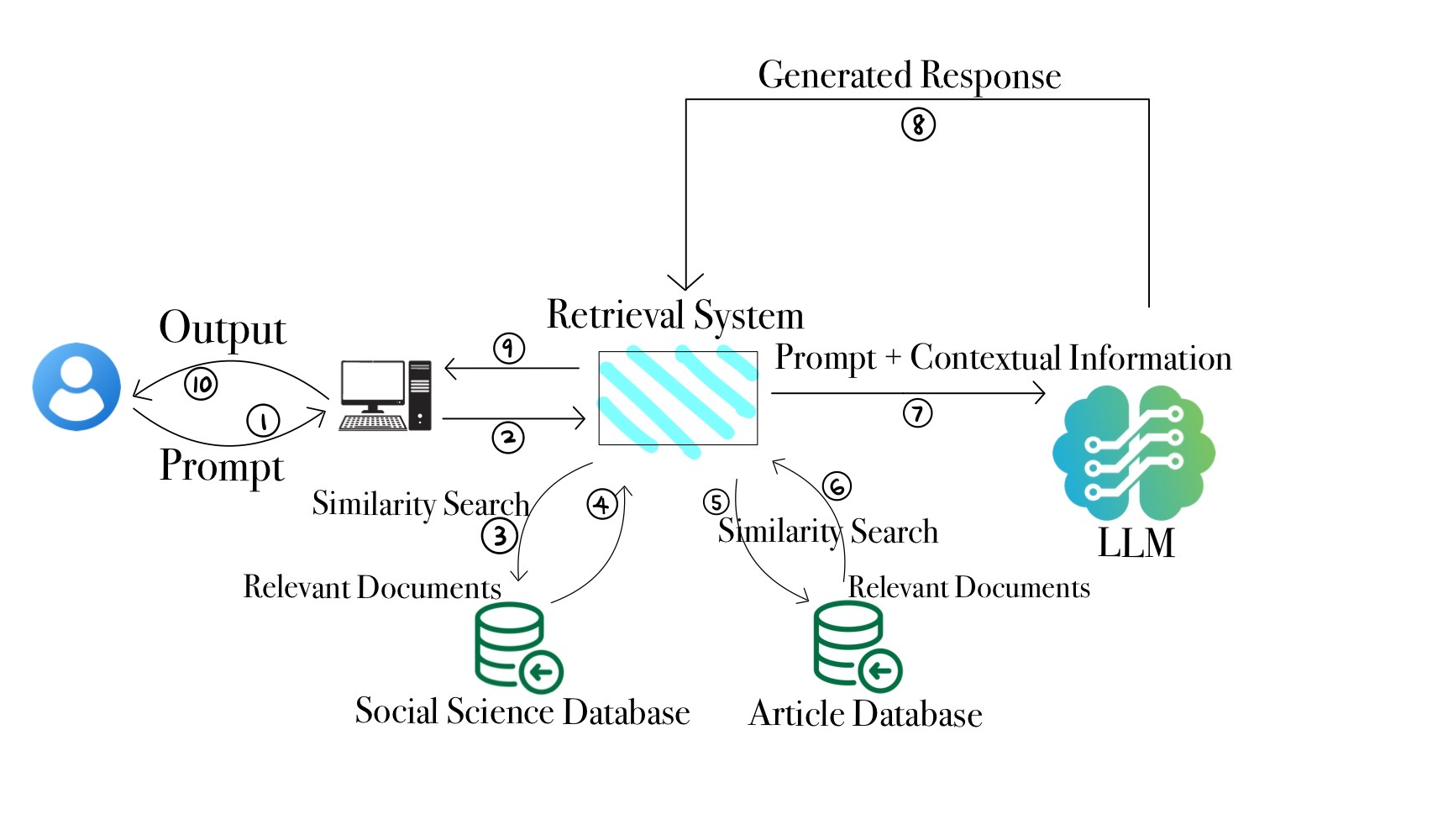}
    \caption{Workflow of the RAG implementation process.}
    \label{fig:rag_workflow}
\end{figure}

\subsection{Identifying PIR and NIR in Countries}

Identifying Positive Intergroup Reciprocity (PIR) and Negative Intergroup Reciprocity (NIR) within media articles provides a deeper understanding of intergroup relations as represented in the media. A total of 6,000 articles per country were analyzed to determine the percentage of PIR-aligned articles out of the total, providing a measure of each country's alignment with intergroup reciprocity and peace. The methodology for this analysis includes the following steps:

\subsubsection{Initial Embedding Search}

Initially, we attempted to identify instances of PIR and NIR by searching for their respective embeddings within the vector database. However, this approach did not yield satisfactory results, as the embedding model struggled to differentiate between positive and negative forms of intergroup reciprocity. Both PIR and NIR were perceived as similar by the model, likely because they both fall under the broader category of "intergroup reciprocity."

\subsubsection{Refining Definitions and Search Criteria}

To address this issue, we refined our definitions and search criteria:

\begin{itemize}
    \item \textbf{PIR:} Defined as "Intergroup tolerance, respect, kindness, help, or support."
    \item \textbf{NIR:} Defined as "Intergroup intolerance, disrespect, aggression, obstruction, or hindrance."
\end{itemize}

With these refined definitions, a new search was conducted within the vector database, yielding more distinct and relevant results. The revised search criteria allowed the model to better capture the nuances between positive and negative reciprocity in intergroup relations.

\subsubsection{Calculating Peace Levels Based on PIR and NIR}

Using the improved search results, we introduced a method to calculate peace levels based on the relative prevalence of PIR and NIR in media articles. This calculation involves determining the percentage of articles that align more closely with PIR than with NIR. By doing so, we provide an alternative but meaningful assessment of each country's peace level, grounded in the representation of intergroup relations in the media.

\section{Results}
The results of our study demonstrate the effectiveness of using a RAG model and refined PIR/NIR definitions to classify peace in global media. By focusing on intergroup reciprocity, we were able to provide a more detailed and contextually relevant analysis of peace as represented in media articles. The following subsections outline the key findings from our analysis.

\subsection{RAG Pipeline Insights}

The RAG pipeline's ability to retrieve and analyze real-world examples of Positive Intergroup Reciprocity (PIR) and Negative Intergroup Reciprocity (NIR) has provided new insights into the dynamics of peace and conflict as portrayed in global media. Below, we discuss some specific outputs from the RAG process that illustrate these concepts in various countries. The results presented in this section stem from the articles retrieved and analyzed by the RAG-modified prompt.

\subsubsection{Example of Positive Intergroup Reciprocity (PIR)}

\textbf{Singapore (SG):} One article retrieved by the RAG pipeline described efforts to move beyond mere tolerance to active "religious appreciation" within the community. The article highlighted instances where people of different faiths participated in each other's religious festivals, such as Muslims and Hindus wishing their Chinese friends well during Chinese New Year and visiting each other during Deepavali and Hari Raya celebrations. This behavior exemplifies PIR, where mutual respect, understanding, and appreciation across different groups contribute to social harmony and peace. The article states: 
\begin{quote}
"...we need to go beyond tolerance and have appreciation, which is boundless. There are many positive signs that we have religious appreciation in Singapore..."
\end{quote}
This real-life example shows how fostering intergroup appreciation and support can strengthen societal bonds and maintain peace.

\subsubsection{Example of Negative Intergroup Reciprocity (NIR)}

\textbf{Kenya (KE):} In contrast, an article from Kenya highlights the impact of political intolerance and aggression on intergroup relations. The report details violent incidents during political rallies where rival groups clashed, leading to injuries and heightened tensions. For example, during a pro-referendum rally, participants from an anti-referendum group attacked and injured attendees, escalating hostility and division between the groups:
\begin{quote}
"In Wote, five people were shot and injured as political rivals in Makueni threw decorum to the wind and instead resorted to violence. In Kapsabet, the Nandi County Governor's chief of staff, who was in a pro-referendum group at an anti-referendum rally, sustained serious head injuries."
\end{quote}
This example illustrates NIR, where aggression and hostility between political groups deteriorate social cohesion and exacerbate conflict, hindering peace efforts.

\begin{table}[htbp]
\centering
\caption{Percentage Closer to PIR by Country}
\begin{tabular}{|l|l|}
\hline
\textbf{Country} & \textbf{Percentage Closer to PIR (Normalized)} \\
\hline
New Zealand & 100\% \\
Singapore & 98\% \\
Ireland & 79\% \\
Australia & 76\% \\
Canada & 76\% \\
Great Britain & 73\% \\
Tanzania & 64\% \\
Sri Lanka & 62\% \\
Bangladesh & 59\% \\
Ghana & 59\% \\
Malaysia & 54\% \\
Jamaica & 36\% \\
Nigeria & 26\% \\
Philippines & 17\% \\
Hong Kong & 12\% \\
India & 11\% \\
US & 2\% \\
Kenya & 0\% \\
\hline
\end{tabular}
\label{table:pir_percentages}
\end{table}

\subsection{Results of the PIR vs NIR Analysis}

The results, as presented in Table \ref{table:pir_percentages}, reveal significant variation in the portrayal of intergroup relations across countries when viewed in a comparative context. Some key findings include:

\begin{itemize}
    \item \textbf{High Comparative PIR Alignment:} Countries such as New Zealand (100\%), Singapore (98\%), and Ireland (79\%) demonstrated the highest comparative alignment with PIR. This suggests that, relative to other countries in the study, the media in these nations predominantly depict intergroup relations positively, focusing more on tolerance, respect, and mutual support. Such portrayals, when viewed comparatively, may contribute to a stronger overall perception of social cohesion and stability.

    \item \textbf{Moderate Comparative PIR Alignment:} Countries like Malaysia (54\%) and Sri Lanka (62\%) exhibited moderate comparative PIR alignment. This indicates that, relative to the entire dataset, the portrayal of intergroup relations in these countries is more balanced, showing both positive and negative aspects. These results might reflect the presence of both supportive and contentious social dynamics within the media narratives.

    \item \textbf{Low Comparative PIR Alignment:} On the lower end of the spectrum, countries such as Hong Kong (12\%), India (11\%), and the United States (2\%) displayed low comparative alignment with PIR. In the context of this analysis, this means that, relative to other countries, the media in these nations are more likely to emphasize negative intergroup relations. Kenya’s 0\% PIR alignment indicates a particularly stark focus on negative intergroup reciprocity in its media coverage, suggesting that media narratives in these regions are more prone to highlighting conflict or discord.

    \item \textbf{Comparative Insights:} The normalized percentages provide a comparative lens, allowing us to see not just the absolute but the relative portrayal of intergroup relations across different media landscapes. For example, the high PIR alignment in New Zealand compared to Kenya's low alignment suggests a significant disparity in how media contribute to the narrative of intergroup relations. This comparative perspective is critical for understanding the broader patterns of peace and conflict as mediated through national media.
\end{itemize}

\section{Discussion}

The results of our study demonstrate the effectiveness of using a RAG model and refined PIR/NIR definitions to classify peace in global media. By focusing on intergroup reciprocity, we were able to provide a more detailed and contextually relevant analysis of peace as represented in media articles. This approach offers several advantages over traditional methods of measuring peace, such as the Positive Peace Index (PPI) and the Human Development Index (HDI) \cite{ppi} \cite{hdi}.

\subsection{Comparing PIR/NIR Analysis with Traditional Peace Measurements}

Traditional peace measurements like the PPI and HDI often rely on quantitative data that may not fully capture the nuances of societal dynamics and intergroup relations. For instance, while the PPI assesses factors such as governance, economic stability, and social structures, it does not directly measure the quality of intergroup relations or the media's portrayal of these dynamics. Similarly, the HDI provides a broad measure of human development but does not account for the subtleties of how peace is constructed or deconstructed through everyday interactions and media representations.

In contrast, the PIR/NIR analysis provides a complementary perspective by focusing on the qualitative aspects of media content that reflect intergroup reciprocity. This method allows for a more nuanced understanding of peace as it is experienced and represented in various cultural contexts. By analyzing how media portrays acts of tolerance, kindness, and support (PIR), versus intolerance, aggression, and hindrance (NIR), we can gain insights into the underlying factors that promote or inhibit peace beyond what is captured by traditional indices.

\subsection{Implications for Peace Measurement and Policy}

The comparative nature of this analysis provides critical insights into how different countries’ media landscapes influence public perceptions of peace and intergroup relations. Countries with higher comparative PIR alignment suggest media environments where positive intergroup dynamics are more prominent, potentially reinforcing social cohesion and peace. Conversely, countries with lower comparative PIR alignment may face challenges related to social fragmentation, as media narratives may contribute to or reflect underlying societal tensions.

These findings highlight the importance of considering the relative portrayal of intergroup relations in media across different contexts. For policymakers and peacebuilders, this comparative analysis offers an opportunity to understand not just the state of media in isolation but how it relates to broader global trends. By focusing on enhancing positive media portrayals in countries with lower PIR alignment, there could be potential for fostering greater social cohesion and mitigating conflict.

Furthermore, the use of PIR/NIR as an alternative or supplementary measure to traditional peace indices provides a more dynamic and real-time tool for assessing peace. This approach can be particularly useful in monitoring and responding to shifts in social dynamics that may not be immediately apparent through conventional data sources.

\subsection{Limitations and Future Directions}

While the PIR/NIR analysis offers valuable insights, it is not without limitations. The reliance on media content means that the analysis is influenced by the biases and editorial decisions of media outlets, which can vary significantly between countries and contexts. Future research could explore integrating other data sources, such as social media or community reports, to triangulate findings and provide a more comprehensive view of intergroup relations.

Additionally, expanding the scope of the analysis to include more diverse cultural and regional contexts could help to refine the PIR/NIR definitions and enhance the robustness of the findings. Developing a real-time dashboard that monitors peace levels based on PIR/NIR in global media could also provide valuable insights for policymakers, researchers, and the public.

We plan to extend the RAG process to analyze other factors that contribute to peace, as highlighted by Fry et al. \cite{fry}. Fry’s work identifies a range of cultural, social, and political factors that influence peace. By applying the RAG methodology to these additional dimensions, we could uncover new insights into how various aspects of peace are represented in media and other textual sources. This expansion would allow for a more holistic understanding of the components that foster or hinder peace, offering a richer framework for peace studies.

\section{Conclusion}

This study highlights the potential of using advanced AI techniques, such as RAG models, to analyze intergroup relations and classify peace in global media. By refining our approach to PIR/NIR, we have provided a more accurate and meaningful analysis of the factors that influence national peace. Unlike traditional peace indices such as the Positive Peace Index (PPI) and Human Development Index (HDI), which provide a broad overview of peace and development, our method focuses on the qualitative nuances of media representation and its impact on public perceptions of peace.

The PIR/NIR framework allows for a deeper understanding of the dynamics that contribute to or detract from peace, providing a valuable tool for researchers, policymakers, and practitioners working in peacebuilding and conflict resolution. By focusing on real-time media content, this approach offers a more dynamic and responsive measure of peace that can inform timely interventions and policy decisions.

Future research could expand on this framework by integrating additional data sources, refining the definitions of PIR and NIR, and developing tools for real-time monitoring and analysis. These advancements could further enhance our ability to understand and promote peace in a complex and rapidly changing world.

\end{document}